\documentclass[journal]{IEEEtran}
\usepackage{amssymb}
\usepackage{amsmath}
\usepackage{graphicx}
\usepackage{cite}
\usepackage{txfonts}
\usepackage{mathrsfs} 
\usepackage{bbm}
\usepackage{fontenc}
\usepackage[binary-units]{siunitx}
\usepackage[caption=false,font=footnotesize]{subfig}
\usepackage{tabularx}
\usepackage{multirow}
\usepackage{diagbox}
\usepackage{xcolor}

\begin{document}
\title{Cell-Free Massive MIMO-OFDM Transmission over\\ Frequency-Selective Fading Channels}

\author{Wei~Jiang,~\IEEEmembership{Senior~Member,~IEEE,}
        and Hans~Dieter~Schotten,~\IEEEmembership{Member,~IEEE}
\thanks{\textit{Corresponding author: Wei Jiang (e-mail: wei.jiang@dfki.de)})}
\thanks{W. Jiang and H. D. Schotten are with German Research Centre for Artificial Intelligence (DFKI), Kaiserslautern, Germany, and are also with the University of Kaiserslautern, Germany.}
}

\markboth{}%
{Jiang \MakeLowercase{\textit{et al.}}: Cell-Free massive MIMO-OFDM System}
\maketitle

\begin{abstract}
This letter presents and analyzes orthogonal frequency-division multiplexing (OFDM)-based multi-carrier transmission for cell-free massive multi-input multi-output (CFmMIMO) over frequency-selective fading channels. Frequency-domain conjugate beamforming, pilot assignment, and user-specific resource allocation are proposed.  CFmMIMO-OFDM  is scalable to serve a massive number of users and is flexible to offer diverse data rates for heterogeneous applications. 
\end{abstract}

\IEEEpeerreviewmaketitle

\section{Introduction}

\IEEEPARstart{C}{ell-free} massive multi-input multi-output (CFmMIMO)  has recently received much attention from both academia and industry \cite{Ref_ngo2017cellfree}. Different aspects of CFmMIMO such as resource allocation \cite{Ref_buzzi2020usercentric}, power control \cite{Ref_nayebi2017precoding}, pilot assignment \cite{Ref_zeng2021pilot}, energy efficiency \cite{Ref_ngo2018total}, backhaul constraint \cite{Ref_masoumi2020performance}, and scalability \cite{Ref_bjornson2020scalable} have been studied. However, previous works merely considered frequency-flat fading (narrow-band) channels.  Most of wireless communications nowadays are broadband with signal bandwidths far wider than the \textit{coherence bandwidth}, leading to frequency selectivity. To the authors' best knowledge, only \cite{Ref_jin2020spectral} discussed CFmMIMO over frequency-selective fading channels. But it applied single-carrier transmission that does not support high data rate due to prohibitively high complexity of signal equalization \cite{Ref_jiang2016ofdm}. Moreover, the conventional CFmMIMO systems generally aim to offer uniform service by maximizing the minimum of per-user rate \cite{Ref_nayebi2017precoding}, whereas it neglects the fact that heterogeneous users have differentiated demands on data throughput. 

To fill this gap, this letter will present and analyze orthogonal frequency-division multiplexing (OFDM)-based multi-carrier transmission for CFmMIMO, coined cell-free massive MIMO-OFDM (CFmMIMO-OFDM),  over frequency-selective fading channels. Contamination-free pilot assignment and channel estimation in the uplink time-frequency grid,  and \textit{frequency-domain} conjugate beamforming in the downlink data transmission are provided. A user-specific resource allocation method that enables the scalability to accommodate massive number of users and the flexibility of offering diverse data rates is proposed.  
Per-user and sum data throughput, considering APs equipped with either single or multiple antennas, are evaluated.   \textbf{Notations}:: Bold lower- and upper-case letters denote vectors and matrices, respectively, while $(\cdot)^*$, $(\cdot)^T$,  and $(\cdot)^H$ express conjugate, transpose, and Hermitian transpose. For ease reference, main mathematical symbols and notation are listed in Table \ref{table_notations}.

\begin{table}[!t]
\renewcommand{\arraystretch}{1.3}
\caption{Definitions of Main Mathematical Symbols.}
\label{table_notations}
\centering
\scriptsize
\begin{tabular}{cl}
\hline
\textbf{Symbol} & \textbf{Definition}     \\  \hline 
$\beta_{mk}$  & Large-scale fading between AP antenna $m$ and user $k$ \\ 
$\mathbf{h}_{mk}$ & Small-scale fading between AP antenna $m$ and user $k$\\  
$\mathbf{g}_{mk}$ &  Channel vector between AP antenna $m$ and user $k$   \\ 
$\mathbf{g}_{mk}^N$ &  $N$-point channel vector by padding zeros at the tail of $\mathbf{g}_{mk}$    \\ 
$\tilde{\mathbf{g}}_{mk}$ &  Frequency-domain channel vector, i.e., the DFT of $\mathbf{g}_{mk}^N$    \\ 
$\tilde{g}_{mk}^r$ and $\hat{g}_{mk}^r$ &  Frequency-domain channel response and its estimate \\ 
$\mathbf{x}_m$ and $\tilde{\mathbf{x}}_m$ &  Time- and frequency-domain transmitted symbol vectors    \\ 
$\mathbf{x}_m^{cp}$ and $\mathbf{y}_k^{cp}$ &  Transmitted and received symbol vectors with cyclic prefix    \\ 
$\mathbf{y}_k$ and $\tilde{\mathbf{y}}_k$ &  Time- and frequency-domain received symbol vectors    \\ 
$\mathscr{U}$ and $\mathscr{U}_s$ & The set of all users and the $s^{th}$ user group \\ 
$\mathbb{B}$ and $\mathbb{B}_s$ & The set of all RBs and the RBs assigned to $\mathscr{U}_s$ \\ 
$K$ and $K_r$ & The number of all users and the number of users over $\mathcal{B}_r$ \\ 
$N$ and $N_{RB}$& The number of subcarriers and the number of RBs \\ \hline
\end{tabular}
\end{table}

\section{System Model}
Consider a geographical area where $M$ randomly distributed access points (APs) are connected to a central processing unit (CPU) via a fronthaul network and serve $K$ users. We first assume that each AP and user is equipped with a single antenna as \cite{Ref_ngo2017cellfree} for simple analysis but will demonstrate its adaptability to multi-antenna APs afterwards. In contrast to the conventional CFmMIMO such as \cite{Ref_ngo2017cellfree, Ref_nayebi2017precoding, Ref_buzzi2020usercentric, Ref_ngo2018total, Ref_jin2020spectral, Ref_zeng2021pilot, Ref_bjornson2020scalable, Ref_masoumi2020performance} that requires $K\ll M$, the number of users in CFmMIMO-OFDM is scalable, ranging from small $K\ll M$ to very large $K\gg M$. Users 
are divided into groups and each group is assigned to different resource blocks (RBs). Thus, the constraint that the number of users is far smaller than $M$ is still satisfied on each RB.  

\begin{figure*}[!tbph]
\centerline{
\subfloat[]{
\includegraphics[width=0.38\textwidth]{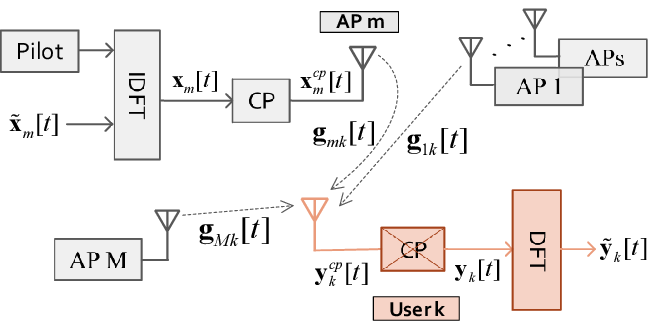}
\label{fig:OFDMTxR}
}
\hspace{20mm}
\subfloat[]{
\includegraphics[width=0.38\textwidth]{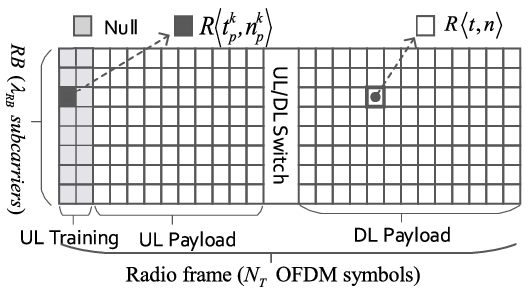}
\label{fig:OFDMGrid}
}
}
\hspace{15mm}
 \caption{Schematic diagram of a CFmMIMO-OFDM system where $M$ AP antennas serve $K$ users: Part (a) elaborates the OFDM transmitter at the $m^{th}$ AP antenna and the receiver at the $k^{th}$ user.  A radio frame consists of $N_T$ OFDM symbols in the axis of time and $N$ subcarriers in the frequency domain, which are grouped into $N_{RB}$ resource blocks.  Part (b) illustrates the time-frequency resource grid of a single RB containing $\lambda_{RB}$ consecutive subcarriers. UL/DL switch denotes the guard interval between the uplink and downlink transmission.}
\label{Fig_Result1}
\end{figure*}

The previous works assume that the small-scale fading is frequency-flat, as modelled by a circularly-symmetric complex Gaussian random variable with zero mean and unit variance, i.e., $h[t] {\sim} \mathcal{CN}(0, 1)$. This assumption is only valid for narrow-band communications.  Nevertheless, most of the current and future mobile communications \cite{Ref_WJ_jiang2021road} are broadband, suffering from severe frequency selectivity. This letter goes beyond  the current state of the art by studying CFmMIMO over a frequency-selective fading channel \cite{Ref_tse2005fundamental}. It can be modeled as a linear time-varying filter $\mathbf{h}[t] {=} \left[ h_0[t],\ldots, h_{L-1}[t]  \right]^T$, where the filter length $L$ should be no less than multi-path delay spread $T_d$ normalized by the sampling interval $T_s$, namely $L\geqslant \left \lceil \frac{T_d}{T_s} \right \rceil  $. 
The tap gain $h_l[t] = \sum_i a_i(tT_s)e^{-j2\pi f_c \tau_i(tT_s)} sinc[l - \tau_i(tT_s)B_w]$ for $l=0,\ldots,L-1$, with  carrier frequency $f_c$, attenuation $a_i(tT_s)$ and delay $\tau_i(tT_s)$ of the $i^{th}$ signal path, signal bandwidth $B_w=1/T_s$, and  $sinc(x)\triangleq \frac{\sin(x)}{x}$ for $x\neq 0$.  The fading channel  between AP $m$ and user $k$ is given by
\begin{equation} \label{Eqn_ChannelModel}
\mathbf{g}_{mk}[t]=\left[ g_{mk,0}[t],\ldots, g_{mk,L_{mk}-1}[t]  \right] ^T=\sqrt{\beta_{mk}[t]} \mathbf{h}_{mk}[t],
\end{equation}
where $g_{mk,l}[t]=\sqrt{\beta_{mk}[t]} h_{mk,l}[t]$  
and $\beta_{mk}[t]$ indicates large-scale fading, which is frequency independent and varies slowly. 

\section{Cell-Free Massive MIMO-OFDM system}
The data transmission in an OFDM system \cite{Ref_jiang2016ofdm} is organized in block-wise, as shown in \figurename \ref{fig:OFDMTxR}.
We write $\tilde{\mathbf{x}}_m[t] = \left[ \tilde{x}_{m,0}[t],\ldots,\tilde{x}_{m,n}[t],\ldots, \tilde{x}_{m,N-1}[t]  \right]^T$ to denote the frequency-domain transmission block of AP $m$ on the $t^{th}$ OFDM symbol. (The tilde $\tilde{}$ marks frequency-domain variables throughout this letter.) Transform $\tilde{\mathbf{x}}_m[t]$ into a time-domain sequence $\mathbf{x}_m[t] = \left[ x_{m,0}[t],\ldots, x_{m,n'}[t],\ldots, x_{m,N-1}[t] \right]^T$ through an $N$-point inverse discrete Fourier transform (IDFT), i.e., $x_{m, n'}[t]=\frac{1}{N}\sum_{n=0}^{N-1}\tilde{x}_{m, n}[t] e^{2\pi jn'n/N}$ for $n'=0,1,\ldots,N-1$ and $j^2=-1$. Defining the discrete Fourier transform (DFT) matrix
\begin{equation}
\label {Eqn_DFTMatrix}
\mathbf{F} =
\left[ \begin{aligned}
         \omega_N^{0\cdot0} &  &  \cdots && \omega_N^{0\cdot(N-1)} \\
         \vdots &&  \ddots && \vdots \\
         \omega_N^{(N-1)\cdot0} &  &  \cdots && \omega_N^{(N-1)\cdot(N-1)}
\end{aligned} \right]
\end{equation}
with a primitive $N^{th}$ root of unity  $\omega_N^{n\cdot n'}=e^{2\pi jnn'/N}$, the OFDM modulation can be written in matrix form as
\begin{equation} \label{eqn:transmitsignal}
   \mathbf{x}_m[t] =\mathbf{F}^{-1} \tilde{\mathbf{x}}_m[t]=\frac{1}{N}\mathbf{F}^{*}\tilde{\mathbf{x}}_m[t]. 
\end{equation}
A guard interval known as cyclic prefix (CP) is added between two consecutive blocks to avoid inter-symbol interference (ISI) and preserve orthogonality of subcarriers. Thus, we get 
$\mathbf{x}_m^{cp}[t]= \left[
               x_{m, N-L_{cp}}[t], \ldots,  x_{m,N-1}[t], x_{m,0}[t], \ldots, x_{m,N-1}[t]\right]^T$ as the transmitted signal.
The ISI can be eliminated if the length of CP  is larger than the length of any channel filter, i.e., $L_{cp} > \max \left( L_{mk}\right)$, for all $m\in\{1,\ldots,M\}$ and $k\in\{1,\ldots,K\}$. 

The signal $\textbf{x}_m^{cp}[t]$ goes through the channel $\mathbf{g}_{mk}[t]$ to reach the typical user $k$, resulting in $ \textbf{x}_m^{cp}[t] \star \mathbf{g}_{mk}[t]$, where $\star$ denotes \textit{the linear convolution}. Thus, the overall received signal at user $k$ is $\mathbf{y}_k^{cp}[t]=\sum_{m=1}^M \textbf{x}_m^{cp}[t] \star \mathbf{g}_{mk}[t]+\mathbf{z}_k[t]$, where $\mathbf{z}_k[t]$ is a vector of additive noise.
Removing the CP, we get
\begin{equation} \label{eqn:reveivedsignal}
  \mathbf{y}_k[t]=\sum_{m=1}^M  \mathbf{g}_{mk}^N[t]\otimes \mathbf{x}_m[t] +\mathbf{z}_k[t],  
\end{equation}
where $\otimes$ stands for \textit{the cyclic convolution} \cite{Ref_jiang2016ofdm} and $\mathbf{g}_{mk}^N[t]$ is an $N$-point channel filter formed by padding  zeros at the tail of $\mathbf{g}_{mk}[t]$, i.e., $\mathbf{g}_{mk}^N[t]{=} \left[ g_{mk,0}[t],\ldots, g_{mk,L_{mk}-1}[t], 0,\ldots,0  \right] ^T$. The DFT demodulator outputs 
\begin{equation} \label{eqn:frequencydomainRx}
    \tilde{\mathbf{y}}_k[t] = \mathbf{F}\mathbf{y}_k[t].
\end{equation}
Substituting (\ref{eqn:transmitsignal}) and (\ref{eqn:reveivedsignal}) into (\ref{eqn:frequencydomainRx}), and applying \textit{the convolution theorem} for DFT \cite{Ref_jiang2016ofdm}, we have
\begin{IEEEeqnarray}{lll}
\label{Eqn_conditionedx} \nonumber
\tilde{\mathbf{y}}_k[t] &=& \sum_{m=1}^{M} \mathbf{F}\left(\mathbf{g}_{mk}^N[t]\otimes\mathbf{x}_m[t]\right)+\mathbf{F}\mathbf{z}_k[t]\\
         & =& \sum_{m=1}^{M} \tilde{\mathbf{g}}_{mk}[t] \odot \tilde{\mathbf{x}}_m[t] + \tilde{\mathbf{z}}_k[t],
\end{IEEEeqnarray}
where $ \odot$ represents the Hadamard product (element-wise multiplication), $\tilde{\mathbf{g}}_{mk}[t]=\mathbf{F}\mathbf{g}_{mk}^N[t]$ and $\tilde{\mathbf{z}}_k[t]=\mathbf{F}\mathbf{z}_k[t]$ denote frequency-domain channel response and noise, respectively. 
At last a frequency-selective channel is transformed into a set of $N$ independent frequency-flat subcarriers. The signal transmission in the downlink on the $n^{th}$ subcarrier is given by
\begin{equation}
\label{Eqn_OFDMDL}
   \tilde{y}_{k,n}[t]=\sum_{m=1}^M \tilde{g}_{mk, n}[t]\tilde{x}_{m,n}[t]+\tilde{z}_{k,n}[t], \:\:k\in\{1,\ldots,K\}, 
\end{equation}
where $\tilde{g}_{mk, n}[t]$ is the $n^{th}$ element of $\tilde{\mathbf{g}}_{mk}[t]$.
Similarly,  the uplink transmission is expressed by 
\begin{equation} \label{eqn:UplinkTransmission}
   \tilde{y}_{m,n}[t]=\sum_{k=1}^K \tilde{g}_{mk,n}[t]\tilde{x}_{k,n}[t]+\tilde{z}_{m,n}[t],\:\:m\in\{1,\ldots,M\}.  
\end{equation}

\section{The communication process} The downlink (DL) transmission from the APs to the users and the uplink (UL)  from  the users to the APs are separated by time-division multiplexing (TDD) with the assumption of perfect channel reciprocity. A radio frame is mainly divided into three phases: UL training, UL payload transmission, and DL payload transmission.  

\subsection{Uplink Training}

We write $\mathscr{R}\langle t,n \rangle$ to denote a resource unit (RU) on the $n^{th}$ subcarrier of the $t^{th}$ OFDM symbol. The time-frequency resource of a radio frame is divided into $N_{RB}$ RBs, each of which contains $\lambda_{RB}=N/N_{RB}$ (assumed to be an integer) consecutive subcarriers. The $r^{th}$ RB is defined as $\mathcal{B}_r \triangleq \left\{ \mathscr{R}\langle t,n \rangle \vert 1\leq t \leq N_T\: \mathrm{and}\: (r-1)\lambda_{RB} \leq n<r\lambda_{RB} \right\}$, for any $r\in \{1,\ldots, N_{RB}\}$, as illustrated in \figurename \ref{fig:OFDMGrid}. 
The transmission of a radio frame in CFmMIMO is carried out within the \textit{coherent time} and the width of one RB is smaller than the \textit{coherence bandwidth}. 
Consequently, this letter simply adopts the \textit{block fading} model where the channel coefficients for all RUs within one RB are assumed to be identical, we have 
\begin{equation} \label{eqn:CSIofRB}
    \tilde{g}_{mk,n}[t] = \tilde{g}_{mk}^r \impliedby \mathscr{R}\langle t,n \rangle \in \mathcal{B}_r. 
\end{equation}

The channel estimation of a conventional CFmMIMO system relies on time-domain pilot sequences, where the maximal number of orthogonal sequences is $\tau_p$ by using $\tau_p$ pilot symbols. If $K\leqslant \tau_p$ as assumed in \cite{Ref_nayebi2017precoding}, pilot contamination can be avoided. Owing to the limitation of the frame length, however, some users need to share the same sequence when $K>\tau_p$, leading to pilot contamination \cite{Ref_zeng2021pilot}. In contrast, CFmMIMO-OFDM is able to provide more orthogonal pilots by means of frequency-division multiplexing thanks to the extra degree of freedom gained from the frequency domain. 
To estimate $\tilde{g}_{mk}^r$,  each user on $\mathcal{B}_r$ needs only one pilot symbol. Suppose the first $\tau_p$ OFDM symbols are dedicated to the uplink training, one RB has $N_p=\tau_p \lambda_{RB}$ orthogonal pilots.
The number of users allocated to $\mathcal{B}_r$ is denoted by $K_r$, there is no pilot contamination if $K_r\leqslant N_p$, which is a very relaxed condition. We write $\mathscr{R} \langle t_p^k,n_p^k \rangle$ with $1 \leqslant t_p^k \leqslant \tau_p$ and $(r-1)\lambda_{RB} \leqslant n_p^k<r\lambda_{RB}$ to denote the RU reserved for the pilot symbol of user $k$, $k\in \{1,\ldots, K_r\}$. Other users keep silence (null) on this RU to achieve orthogonality. As illustrated in \figurename \ref{fig:OFDMGrid}, the pilot assignment is  specified by
\begin{equation}
\label{eqn:PilotAssignment} 
\tilde{x}_{k,n}[t]=
    \begin{cases}
    \sqrt{p_u}\: \mathbb{P}_k,& \text{if } t=t_p^k \land  n=n_p^k\\
    0,              & \text{otherwise}
\end{cases}, \:\:1\leq t\leq \tau_p,
\end{equation}
where $\land$ represents logical $\mathrm{AND}$, $\mathbb{P}_k$ is the known pilot symbol with $\mathbb{E}[\vert \mathbb{P}_k\vert^2]=1$, and $p_u$ denotes the UL transmit power limit.

Substituting (\ref{eqn:CSIofRB}) and (\ref{eqn:PilotAssignment})  into (\ref{eqn:UplinkTransmission}), we get the received signal of the $m^{th}$ AP on $\mathscr{R} \langle t_p^k,n_p^k \rangle$  as 
\begin{IEEEeqnarray}{lll} \nonumber \label{eqn:PilotRx}
  \tilde{y}_{m,n_p^k}[t_p^k] &=&\sum_{k=1}^{K_r} \tilde{g}_{mk,n_p^k}[t_p^k]\tilde{x}_{k,n_p^k}[t_p^k]+\tilde{z}_{m,n_p^k}[t_p^k] \\ \nonumber
  &=& \tilde{g}_{mk,n_p^k}[t_p^k]\tilde{x}_{k,n_p^k}[t_p^k]+\sum_{k'\neq k}^{K_r} \tilde{g}_{mk',n_p^{k}}[t_p^k]\tilde{x}_{k',n_p^k}[t_p^{k}]+\tilde{z}_{m,n_p^k}[t_p^k]\\ \nonumber
  &=& \sqrt{p_u}\tilde{g}_{mk,n_p^k}[t_p^k]\mathbb{P}_k+\tilde{z}_{m,n_p^k}[t_p^k]\\
  &=& \sqrt{p_u}\tilde{g}_{mk}^r\mathbb{P}_k+\tilde{z}_{m,n_p^k}[t_p^k].
\end{IEEEeqnarray}
Let $\hat{g}_{mk}^r$ be an estimate of $\tilde{g}_{mk}^r$, we have  $\hat{g}_{mk}^r = \tilde{g}_{mk}^r-\xi_{mk}^r$ with estimation error $\xi_{mk}^r$ raised by additive noise.
Applying the minimum mean-square error (MMSE) estimation gets 
\begin{equation} \label{eqn:MMSE}
    \hat{g}_{mk}^r = \left(\frac{R_{gg}\mathbb{P}_k^*}{R_{gg}|\mathbb{P}_k|^2 + R_{nn}}\right)\tilde{y}_{m,n_p^k}[t_p^k]=\left(\frac{\beta_{mk}\mathbb{P}_k^*}{\beta_{mk}|\mathbb{P}_k|^2 + \sigma_z^2}\right)\tilde{y}_{m,n_p^k}[t_p^k],
\end{equation}
which applies $R_{gg}=\mathbb{E}\left[ \left \vert \tilde{g}_{mk}^r\right \vert^2\right]=\beta_{mk}$ and $R_{nn}=\mathbb{E}\left[ \left \vert \tilde{z}_{m,n}[t]\right \vert^2\right]=\sigma_z^2$.  Compute the variance of $\hat{g}_{mk}^r$ as
\begin{IEEEeqnarray}{lll} \nonumber
\mathbb{E}\left[\hat{g}_{mk}^r{(\hat{g}_{mk}^r)}^*\right]&=&\mathbb{E}\left[\frac{\beta_{mk}^2 \left \vert \mathbb{P}_k \right \vert ^2 \left \vert \sqrt{p_u} \tilde{g}_{mk}^r\mathbb{P}_k+\tilde{z}_{m,n_p^k}[t_p^k]\right \vert ^2}{(\beta_{mk}|\mathbb{P}_k|^2 + \sigma_z^2)^2}\right] \\ \nonumber &=&\frac{\beta_{mk}^2\mathbb{E}\left[\left \vert \sqrt{p_u} \tilde{g}_{mk}^r\mathbb{P}_k+\tilde{z}_{m,n_p^k}[t_p^k] \right \vert^2\right]}{(\beta_{mk} + \sigma_z^2)^2}=\frac{p_u\beta_{mk}^2}{p_u\beta_{mk} + \sigma_z^2}. 
\end{IEEEeqnarray}
Consequently, we know that $\hat{g}_{mk}^r\in \mathcal{CN}(0,\alpha_{mk})$ with $\alpha_{mk}=\frac{p_u\beta_{mk}^2}{p_u\beta_{mk} + \sigma_z^2}$, in comparison with $\tilde{g}_{mk}^r\in \mathcal{CN}(0,\beta_{mk})$.

\subsection{Uplink Payload Data Transmission}
Suppose $\tau_u$ OFDM symbols are used for uplink transmission, on the RU $\mathscr{R}\langle t,n \rangle \in \mathcal{B}_r$, $\tau_p<t\leq\tau_p+\tau_u$, 
all $K_r$ users simultaneously transmit their signals to the APs. The $k^{th}$ user weights its transmit symbol $q_{k,n}[t]$, $\mathbb{E}\left[  \vert q_{k,n}[t] \vert^2 \right ]=1$, by a power-control coefficient $\sqrt{\psi_{k}}$, $0\leq \psi_{k}\leq 1$. Substituting $\tilde{x}_{k,n}[t]=\sqrt{\psi_{k}p_u}q_{k,n}[t]$ into (\ref{eqn:UplinkTransmission}) yields 
\begin{equation}
   \tilde{y}_{m,n}[t]=\sqrt{p_u} \sum_{k=1}^{K_r} \tilde{g}_{mk,n}[t]\sqrt{\psi_{k}}q_{k,n}[t]+\tilde{z}_{m,n}[t]. 
\end{equation}

\subsection{Downlink Payload Data Transmission}

Inspired by \cite{Ref_ngo2017cellfree, Ref_nayebi2017precoding, Ref_zeng2021pilot, Ref_buzzi2020usercentric, Ref_ngo2018total} that apply conjugate beamforming in the downlink, CFmMIMO-OFDM employs  conjugate beamforming in the frequency domain. On $\mathscr{R}\langle t,n \rangle\in \mathcal{B}_r$, $\tau_p+\tau_u< t \leq N_T$, each AP multiplexes a total of $K_r$ symbols, i.e., $s_{k,n}[t]$ intended to user $k$, $k=1,\ldots,K_r$, before transmission. With a power-control coefficient $\sqrt{\eta_{mk}}$, $0\leq \eta_{mk}\leq 1$, the transmitted signal of the $m^{th}$ AP is
\begin{equation}
\label{Eqn_ConjugateBFTranmitSymbol}
    \tilde{x}_{m,n}[t] = \sqrt{p_d} \sum_{k=1}^{K_r} \sqrt{\eta_{mk}} {\left(\hat{g}_{mk,n}[t]\right)}^* s_{k,n}[t].
\end{equation}
Substituting (\ref{Eqn_ConjugateBFTranmitSymbol}) into (\ref{Eqn_OFDMDL}) to  get the received signal at user $k$
\begin{IEEEeqnarray}{lll}
\label{Eqn_OFDMdownlink} \nonumber
   \tilde{y}_{k,n}[t] &=&\sqrt{p_d} \sum_{m=1}^M \tilde{g}_{mk,n}[t] \sum_{k'=1}^{K_r} \sqrt{\eta_{mk'}}  {\left(\hat{g}_{mk',n}[t]\right)}^* s_{k',n}[t]+\tilde{z}_{k,n}[t] \\ \nonumber
   &=&\underbrace{\sqrt{p_d} \sum_{m=1}^M \sqrt{\eta_{mk}} \left|\hat{g}_{mk,n}[t]\right|^2 s_{k,n}[t]}_{Desired\:Signal}  \\ &+&\underbrace{\sqrt{p_d}\sum_{m=1}^M \hat{g}_{mk,n}[t] \sum_{k'\neq k}^{K_r}  \sqrt{\eta_{mk'}}  {\left(\hat{g}_{mk',n}[t]\right)}^* s_{k',n}[t] }_{Multi-User\: Interference}\\ \nonumber &+&\underbrace{\sqrt{p_d}\sum_{m=1}^M \xi_{mk,n}[t] \sum_{k'= 1}^{K_r}  \sqrt{\eta_{mk'}}  {\left(\hat{g}_{mk',n}[t]\right)}^* s_{k',n}[t]}_{Channel-Estimate\:Error}+\underbrace{\tilde{z}_{k,n}[t]}_{Noise}.
\end{IEEEeqnarray}
Each user is assumed to have the knowledge of channel statistics $ \mathbb{E} \left [\sum_{m=1}^M \sqrt{\eta_{mk}} \left \vert \hat{g}_{mk,n}[t]\right \vert ^2\right]$ rather than channel realizations $\hat{g}_{mk,n}[t]$ since there is no pilot and channel estimation in the downlink. Applying a similar method used in \textit{Theorem 1}  of \cite{Ref_ngo2017cellfree}, we can derive that the spectral efficiency of user $k$ on $\mathscr{R}\langle t,n \rangle \in \mathcal{B}_r$  is lower bounded by $\log_2\left(1+\gamma_{k}^{\langle t,n\rangle}\right)$ with
\begin{equation} \label{eqn:SNR_SAAP}
    \gamma_{k}^{\langle t,n\rangle}=  \frac{p_d \left(\sum_{m=1}^M \sqrt{\eta_{mk}} \alpha_{mk}  \right)^2}
    {\sigma^2_z+p_d \sum_{m=1}^M \beta_{mk} \sum_{k'=1}^{K_r}  \eta_{mk'} \alpha_{mk'} },
\end{equation}
implying that the effect of small-scale fading is vanished (a.k.a. channel hardening), 
such that $\gamma_{k}^{\langle t,n\rangle}=\gamma_{k}^r$, for all $\mathscr{R}\langle t,n \rangle \in \mathcal{B}_r$.

\begin{table*}[!t]
\renewcommand{\arraystretch}{1.3}
\scriptsize
\caption{$95\%$-likely and $50\%$-likely (median) per-user data rates, and averaged sum data throughput for MBB and MTC.}
\label{table_datarates}
\centering
\begin{tabular}{|c|l|l|l|l|l|l|l|l|l|l|l|}
\hline
\multirow{3}{*}{\backslashbox{Rate}{System}} &\multicolumn{7}{|c|}{$M=128$}& \multicolumn{2}{|c|}{$M=256$}&\multicolumn{2}{|c|}{$M=512$}\\ \cline{2-12}
&\multicolumn{4}{|c|}{MBB}&\multicolumn{3}{|c|}{MTC}&MBB&MTC&MBB&MTC\\ \cline{2-12}
&$K=6$& $K=12$ & $K=24$ & $K=36$ &$K=1200$&$K=2400$&$K=3600$ &$K=6$&$K=3600$ &$K=6$&$K=3600$\\ \hline
$95\%$-likely [Mbps] &28.20&19.96&10.34&5.66&0.199&0.103&0.057& 40.47 &0.137&54.40&0.229 \\ \hline
Median [Mbps] & 46.45 &38.19&27.64&20.61&0.381&0.276&0.206&59.55 & 0.297 &74.16&0.401\\ \hline
Sum [Mbps] &273.70 &448.21&644.92&720.70&447.83&644.50&721.23&351.57&1034.97&437.44&1397.28 \\ \hline
\end{tabular}
\end{table*}

\section{User-specific Resource Allocation}

The conventional CFmMIMO systems \cite{Ref_ngo2017cellfree, Ref_nayebi2017precoding, Ref_buzzi2020usercentric, Ref_ngo2018total, Ref_jin2020spectral, Ref_zeng2021pilot, Ref_bjornson2020scalable, Ref_masoumi2020performance} can support only very few $K\ll M$ users  with uniform quality of service. By exploiting the frequency domain, CFmMIMO-OFDM is adaptive to  different numbers of users from a few $K \ll M$ to massive $K\gg M$ and is flexible to offer diverse data rates for heterogeneous users.  Classify all users $\mathscr{U}=\{u_1,u_2,\cdots,u_K\}$ into  different groups $\mathscr{U}_s$, $s=1,2,\ldots,S$ in terms of their demands on data throughput, subjecting to $\bigcup_{s=1}^{S}\mathscr{U}_s=\mathscr{U}$ and $\mathscr{U}_s \bigcap \mathscr{U}_{s'}=\varnothing$, $\forall s'\neq s$.  The number of users in $\mathscr{U}_s$  satisfies $\left\vert \mathscr{U}_s \right\vert\ll M$ and $\sum_{s=1}^S \left\vert \mathscr{U}_s \right\vert=K$, where $\left\vert \cdot \right\vert$ stands for the cardinality of a set.
The resource pool is $\mathbb{B}=\left\{ \mathcal{B}_r \vert 1\leq r \leq N_{RB}\right\}$ where the granularity for allocation is one RB.  Using $\mathbb{B}_s$ to denote the RBs allocated to $\mathscr{U}_s$, we have $\bigcup_{s=1}^S \mathbb{B}_s \in \mathbb{B}$ (when all RBs are used, $\bigcup_{s=1}^S \mathbb{B}_s = \mathbb{B}$) and $\mathbb{B}_s \bigcap \mathbb{B}_{s'}=\varnothing$, $\forall s'\neq s$.    
If $\mathcal{B}_r \in \mathbb{B}_s$, the number of users served by this RB is $K_r=\left\vert \mathscr{U}_s \right\vert$.
For any user $k\in \mathscr{U}_s$, its per-user data rate in the downlink is
\begin{equation}
    R_{k}= \left( 1-\frac{\tau_p+\tau_u}{N_T} \right)   \sum_{\mathcal{B}_r  \in \mathbb{B}_s}   \lambda_{RB} \triangle f \log_2\left(1+\gamma_{k}^r\right),
\end{equation}
where $\triangle f$ is the subcarrier spacing. Then, the downlink sum data throughput of the system is $R_d= \sum_{s=1}^S \sum_{k\in \mathscr{U}_s} R_{k}$.

\section{Adaptability to Multi-Antenna AP}
Inspired by \cite{Ref_chen2018channel}, which argues that channel hardening and favorable propagation can be more easily obtained by deploying multiple antennas per AP, we further study CFmMIMO-OFDM with multi-antenna APs \cite{Ref_ngo2018total}. Generalizing the number of antennas per AP to $N_t\geqslant 1$, there are $N_{AP}=M/N_t$ (assumed to be an integer) homogeneous APs indexed by $q=1,\ldots,N_{AP}$.  The only difference raised by multi-antenna APs is that collocated antennas have identical large-scale fading, 
i.e., $\beta_{mk}=\beta_{q k}$ for any $m=(q-1)N_t+1,\ldots,qN_t$. As \cite{Ref_ngo2018total} and \cite{Ref_chen2018channel}, small-scale fading from each collocated antenna to the typical user is assumed to be independent. As a result, the channel model with multi-antenna APs is simply a special case of $\mathbf{g}_{mk}$ in (\ref{Eqn_ChannelModel}), which is transparent to the communication process and does not need any modifications. When the users transmit pilot symbols, see (\ref{eqn:PilotRx}), the observation at AP $q$ is 
\begin{equation} \nonumber \label{eqn:PilotRx_MultiAntennaAP}
  \tilde{\mathbf{y}}_{q,n_p^k}[t_p^k] = \sqrt{p_u}\tilde{\mathbf{g}}_{qk}^r\mathbb{P}_k+\tilde{\mathbf{z}}_{q,n_p^k}[t_p^k],
\end{equation}
where $\tilde{\mathbf{y}}_{q}$ and $\tilde{\mathbf{z}}_{q}$ are $N_t\times 1$ received symbol and noise vectors, respectively, and $\tilde{\mathbf{g}}_{qk}^r=\left [\tilde{g}_{(q-1)N_t+1,k}^r,\ldots,\tilde{g}_{qN_t,k}^r\right]^T \in \mathcal{C}^{N_t\times 1}$ stands for the channel vector between AP $q$ and user $k$. Similar to (\ref{eqn:MMSE}), the MMSE estimate of $\tilde{\mathbf{g}}_{qk}^r$ can be obtained by
\begin{equation} \label{eqn:MMSE_MAAP}
    \hat{\mathbf{g}}_{qk}^r = \left(\frac{\mathbf{R}_{gg}\mathbb{P}_k^*}{\mathbf{R}_{gg}|\mathbb{P}_k|^2 + \mathbf{R}_{nn}}\right)\tilde{\mathbf{y}}_{q,n_p^k}[t_p^k]=\left(\frac{\beta_{qk}\mathbb{P}_k^*}{\beta_{qk}|\mathbb{P}_k|^2 + \sigma_z^2}\right)\tilde{\mathbf{y}}_{q,n_p^k}[t_p^k],
\end{equation}
which applies $\mathbf{R}_{gg}=\beta_{qk}\mathbf{I}_{N_t}$ and $\mathbf{R}_{nn}=\sigma_z^2\mathbf{I}_{N_t}$ and we get  $\hat{\mathbf{g}}_{qk}^r\in \mathcal{CN}(\mathbf{0},\alpha_{qk}\mathbf{I}_{N_t})$ with $\alpha_{qk}=\frac{p_u\beta_{qk}^2}{p_u\beta_{qk} + \sigma_z^2}$.
The transmitted signal can also be expressed by (\ref{Eqn_ConjugateBFTranmitSymbol}) since each antenna independently generates its signal using local channel estimates,  regardless of how many antennas per AP.  With conjugate beamforming, the power control  is generally decided by $\beta_{qk}$ \cite{Ref_nayebi2017precoding}, we have $\eta_{mk}=\eta_{qk}$ for $m=(q-1)N_t+1,\ldots,qN_t$. Then, (\ref{eqn:SNR_SAAP}) can be rewritten for the case of multi-antenna AP as
\begin{equation} \label{eqn:SNR_MAAP}
    \gamma_{k}^{\langle t,n\rangle}=  \frac{p_d N_t^2\left(\sum_{q=1}^{N_{AP}} \sqrt{\eta_{qk}} \alpha_{qk}  \right)^2}
    {\sigma^2_z+p_d N_t \sum_{q=1}^{N_{AP}} \beta_{qk} \sum_{k'=1}^{K_r}  \eta_{qk'} \alpha_{qk'} }.
\end{equation}

\begin{figure*}[!tbph]
\centerline{
\subfloat[]{
\includegraphics[width=0.39\textwidth]{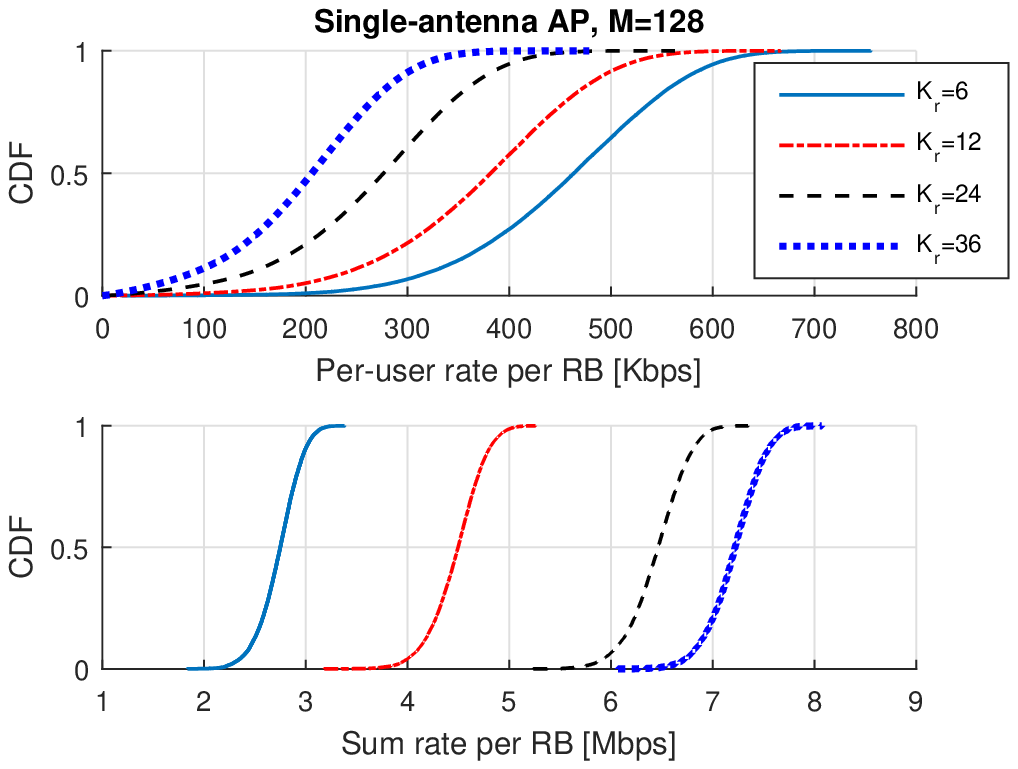}
\label{fig:result1}
}
\hspace{20mm}
\subfloat[]{
\includegraphics[width=0.39\textwidth]{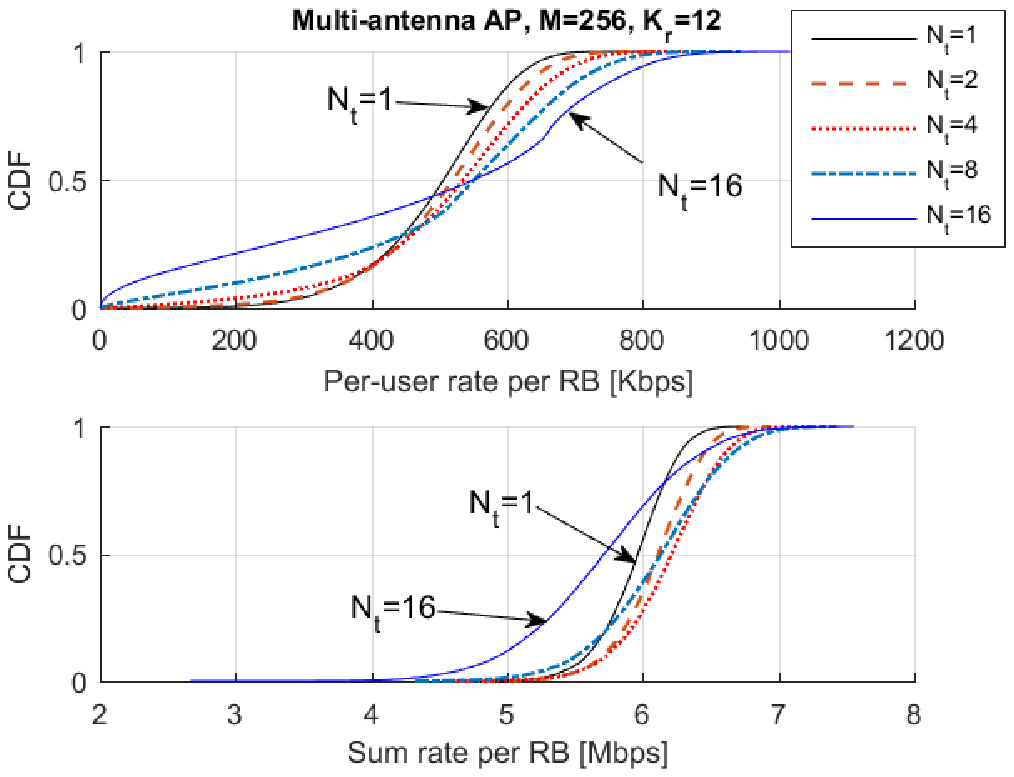}
\label{fig:result2}
}
}
\hspace{15mm}
 \caption{Performance comparison with respect to the CDFs of per-user and sum data rates over a single RB.  Part (a) utilizes $M=128$ \textit{single-antenna} APs and the number of users varies from $K_r=6$, $12$, $24$, to $36$. In part (b), a total of $M=256$ antennas are distributed over a set of \textit{multi-antenna} APs (with $N_t=1$, $2$, $4$, $8$ or $16$) to serve $K_r=12$ users.  }
\label{Fig_performance}
\end{figure*}

\section{Numerical Results}
The performance of CFmMIMO-OFDM in terms of per-user and sum rates is evaluated. 
We consider a square area of $1\times 1\mathrm{km^2}$ where a total of $M$ AP antennas serve $K$ users. The COST-Hata model in \cite{Ref_ngo2017cellfree} is applied for the path loss, 
where  
the three-slope breakpoints  take values $d_0=10\mathrm{m}$ and $d_1=50\mathrm{m}$ while $L=140.72\mathrm{dB}$ with carrier frequency $f_c=1.9\mathrm{GHz}$, the height of AP antenna $h_{AP}=15\mathrm{m}$, and the height of user antenna $h_u=1.65\mathrm{m}$. 
Use $\beta_{mk}=10^\frac{PL_{mk}+X_{mk}}{10}$ to get large-scale fading, where the shadowing fading $X_{mk}\sim \mathcal{N}(0,\sigma_{sd}^2)$ with $\sigma_{sd}=8\mathrm{dB}$. 3GPP Extended Typical Urban (ETU) model specified by delay profile in $\mathrm{millisecond}$ of $\{0,	0.05,	0.12,	0.2,	0.23,	0.5,	1.6,	2.3,	5\}$ with  relative power as $\{-1,	-1,	-1,	0,	0,	0,	-3,	-5,	-7\}$ is applied for small-scale fading\footnote{The near-far effect raised by distributed APs not only generates different distance-dependent large-scale fading but also causes diverse delay spreads in small-scale fading. Due to channel hardening, however, small-scale fading does not affect the final performance of conjugate beamforming, which relies only on large-scale fading. This argument can be proved by (\ref{eqn:SNR_SAAP}) and (\ref{eqn:SNR_MAAP}), and is observed also in simulations. However, it is necessary to  consider heterogeneous delay spreads raised by cell free anywhere small-scale fading can affect performance. }.
The maximum transmit power of AP antenna and user are $p_d=0.2\mathrm{W}$ and $p_u=0.1\mathrm{W}$, respectively. The APs simply adopt the full-power transmission strategy, i.e.,  $\eta_{mk}=\left(\sum_{k=1}^{K_r} \alpha_{mk} \right)^{-1}$.  The white noise power density  is $-174\mathrm{dBm/Hz}$ with a noise figure of $9\mathrm{dB}$. Deducting a guard band of $1\mathrm{MHz}$ at both sides, the signal bandwidth of $B_w=20\mathrm{MHz}$ is divided into $N=1200$ subcarriers with $\triangle f=15\mathrm{kHz}$. Each RB consists of $\lambda_{RB}=12$ subcarriers and thus $N_{RB}=100$. We apply $N_T=10$ for a frame including $\tau_p$ pilot symbols and $N_T-\tau_p$ downlink symbols, neglecting uplink transmission (i.e., $\tau_u=0$) in our simulation. 

\figurename \ref{fig:result1} compares cumulative distribution functions (CDFs) of per-user and sum rates over a single RB (i.e., $180\mathrm{kHz}$) achieved by CFmMIMO-OFDM with $M=128$ single-antenna APs and $K_r=6$, $12$, $24$, or $36$.  
As we expected, per-user rate decreases when more users share the same amount of resource.  But the more users are served together, the higher sum data throughput is achieved. In contrast to a single RB, the system can offer diverse transmission rates through the user-specific resource allocation. Table \ref{table_datarates} shows the achievable capacity of the whole system using $100$ RBs.  We are interested in two extreme cases: mobile broadband (MBB) that requires very high rate and machine-type communications (MTC) with a massive number of low-rate users. Let $K=6$ MBB users (only one group) consume all $100$ RBs, the $95\%$-likely per-user rate can reach $28.2\mathrm{Mbps}$ and the $50\%$-likely or median rate is $46.45\mathrm{Mbps}$ with $M=128$. Increased to $M=512$, an MBB user can get a $95\%$-likely rate of $54.4\mathrm{Mbps}$ and a median rate of $74.16\mathrm{Mbps}$. In contrast,  we equally divide $K=3600$ users into $S=100$ groups  to accommodate as many MTC users as possible.   Each group with $K_r=36$ users is allocated with a single RB. With $M=128$, each MTC user gets a $95\%$-likely rate of $57\mathrm{Kbps}$ and a median rate of $206\mathrm{Kbps}$. Using $M=512$, the system can support $3600$ MTC users with an experienced rate of $229\mathrm{Kbps}$. In addition, the performance of multi-antenna APs is also evaluated where a total of $M=256$ antennas serve $K_r=12$ users. When the number of antennas per AP increases from $N_t=1$ to $2$, both per-user and sum rates boost, as shown in \figurename \ref{fig:result2}. Increased to $N_t=4$, although the sum rate can grow slightly, the $95\%$-likely per-user rate drops. Remarkable performance loss on both per-user and sum rates is observed if $N_t=16$. That is because some users may locate far away from any APs when there are only a few APs, implying the Marco-diversity stemming from CFmMIMO vanishes. 

\section{Conclusion}
The proposed CFmMIMO-OFDM system is effective to combat frequency selectivity  and pilot contamination can be eliminated by exploiting frequency-domain orthogonal pilots. In contrast to CFmMIMO that supports only very few users with uniform service quality, it is scalable to the scale of users and is adaptive to diverse service demands. It was verified by numerical results that CFmMIMO-OFDM can accommodate different numbers of users ranging from a few to thousands and can offer data rates from tens of $\mathrm{Kbps}$ to tens of $\mathrm{Mbps}$.

\bibliographystyle{IEEEtran}
\bibliography{IEEEabrv,Ref_COML}

\end{document}